\documentclass[aps,showpacs,twocolumn]{revtex4}
\usepackage{graphicx}% Include figure files
\usepackage{soul}
\usepackage{color}
\usepackage{amsmath}

\begin{document}
\title{Shuttling heat across 1D homogenous nonlinear lattices with a Brownian heat motor}
\author{Nianbei Li$^1$}
\email{Nianbei.Li@Physik.Uni-Augsburg.DE}
\author{Fei Zhan$^1$}
\email{Fei.Zhan@Physik.Uni-Augsburg.DE}
\author{Peter H\"anggi$^{1,2}$}
\email{Hanggi@Physik.Uni-Augsburg.DE}
\author{Baowen Li$^{2,3}$}
\email{phylibw@nus.edu.sg} %\altaffiliation {Corresponding author}
\address{$^1$ Institut f\"ur Physik, Universit\"at Augsburg Universit\"atsstr. 1
        D-86135 Augsburg, Germany\\
         $^2$ Department of Physics and Centre for Computational Science
         and Engineering, National University of Singapore, Republic of Singapore
         117542\\
         $^3$ NUS Graduate School for Integrative Sciences and Engineering,
         Singapore 117597, Republic of Singapore}

%\date{4 March, 2008}

\begin{abstract}
We investigate directed  thermal heat flux across 1D homogenous
nonlinear lattices when no  net thermal bias is present on average.
A nonlinear lattice of Fermi-Pasta-Ulam-type or Lennard-Jones-type
system is connected at both ends to thermal baths which are held at
the same temperature on temporal average. We study two different
modulations of the heat bath temperatures, namely: (i) a symmetric,
harmonic ac-driving of temperature of one heat bath only and (ii) a
harmonic mixing drive of temperature acting on both heat baths.
While for case (i) an adiabatic result for the net heat transport
can be derived in terms of the temperature dependent heat
conductivity of the nonlinear lattice a similar such transport
approach fails for the harmonic mixing case (ii). Then, for case
(ii), not even the sign of the resulting  Brownian motion induced
heat flux can be predicted a priori. A non-vanishing heat flux
(including a non-adiabatic reversal of flux) is detected which is
the result of an induced dynamical symmetry breaking mechanism in
conjunction with the nonlinearity of the lattice dynamics. Computer
simulations demonstrate that the heat flux is robust against an
increase of lattice sizes. The observed ratchet effect for such
directed heat currents is quite sizable for our studied class of
homogenous nonlinear lattice structures, thereby making this setup
accessible for experimental implementation and verification.
\end{abstract}

% 07.20.Pe Heat engines; heat pumps; heat pipes
% 44.90.+c Other topics in heat transfer
%85.90.+h Other topics in electronic and magnetic devices and microelectronics
%05.90.+m Other topics in statistical physics, thermodynamics, and
%nonlinear dynamical systems (restricted to new topics in section
%05)
%05.40.-a   Fluctuation phenomena, random processes, noise, and Brownian motion

\maketitle
\section{Introduction}
In recent years, we have experienced a wealth of theoretical and
experimental activities in the field of phononics, the science and
engineering of phonons \cite{WangLi08}. Traditionally being regarded
as a nuisance, phonons are found to be able to carry and process
information as well as electrons do. The control and manipulation of
phonons manifest itself in the form of theoretically designed
thermal device models such as thermal diodes
\cite{rectifier,diode,quantumdiode,Hu06,peyrard,yang}, thermal
transistors \cite{transistor}, thermal logic gates \cite{WangLi07}
and thermal memories \cite{wangli08}. The theoretical research has
been accompanied by pioneering experimental efforts. In particular,
the first realization of solid-state thermal diode has been put
forward with help of asymmetric nanotubes \cite{experimentaldiode}.
Owing to the transport of phonons, the heat flow can be controlled
the same way as electric currents.

Dwelling on ideas from the field of Brownian motors
\cite{BM1,BM2,BM3,BM4,BM5}, -- originally devised for Brownian
particle transport, -- a thermal ratchet based on a nonlinear lattice setup has been proposed in Ref.
\cite{LHL}. In absence of any stationary non-equilibrium bias, a
non-vanishing net heat flow can be induced by  non-biased,
temporally alternating bath temperatures combined with a
non-homogenous coupled nonlinear lattice structure. In the similar
spirit of pumping heat on the molecular scale
\cite{vandenbroeckPRL2006,segal2003,segalPRE2006,marathePRE2007,vandenbroeckPRL2008},
the heat flow can be directed against an external thermal bias.

%Figure 1
\begin{figure}
\includegraphics[width=\columnwidth]{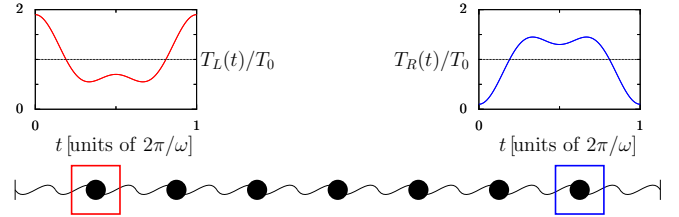}
\vspace{-.5cm} \caption{(Color-online) Schematic setup of a 1D
homogenous nonlinear lattice, being coupled to two heat baths with
periodically varying temperatures at $T_L(t)$ and $T_R(t)$.}
\label{fig:1}
\end{figure}

With this work we propose a superior, easy to implement Brownian
heat motor  that can induce finite net heat flow for a {\it
homogenous}, intrinsically symmetric  lattice structure as depicted
in Fig. \ref{fig:1}. In doing so, the lattice system is brought into
contact with two heat baths which both may be subjected to
time-periodic temperatures, see Fig. \ref{fig:1}. To obtain a finite
directed heat current then requires a symmetry breaking. In this
work we shall investigate two mechanisms of symmetry breaking by use
of a temporal modulation of temperatures of heat baths.

Section II introduces our model for directing heat current through
1D lattice chains with temperature modulations  applied to
connecting  heat baths. Section III presents analytical adiabatic
theory and extensive  numerical results for the case that
temperature is time-modulated in one bath only, i.e. our case (i).
In this case,  an intrinsic nonlinear temperature dependence of the
heat conductivity is sufficient to induce a shuttling of heat. Our
main results are presented with Section IV: A more intriguing
mechanism comes into play  when applying an unbiased modulation of
temperatures in both heat baths. Put differently, in order to
eliminate the possibility of solely creating a non-vanishing net
heat flux from the the temperature-dependent thermal conductivity
$\kappa(T)$ we invoke a nonlinear  harmonic mixing drive of
temperature in both heat baths; i.e. our case (ii). The results are
discussed and summarized in the Conclusions.

\section{Shuttling heat despite vanishing average temperature bias}

Explicitly, we study numerically a 1D homogenous nonlinear lattice
consisting of $N$ atoms of identical mass $m$. This setup is
depicted in Fig. \ref{fig:1}, with the following nonlinear lattice Hamiltonian:
\begin{eqnarray}\label{nonlinear lattice}
H=\sum^{N}_{i=1}\frac{p^2_i}{2m}+\sum^{N}_{i=0}V(x_{i+1},x_{i})
\end{eqnarray}

where $x_i$ is the coordinate for $i$-th atom, and the distance
between two neighboring atoms at equilibrium gives the lattice
constant $a$. The interaction between nearest neighbors is described
by $V(x_{i+1},x_i)$. Here, fixed boundary conditions $x_{0}=0$ and
$x_{N+1}=(N+1)a$ have been employed. The first and last atom are put
into contact with two Langevin heat baths, generally possessing time-dependent
temperatures $T_L(t)$ and $T_R(t)$, respectively.
Moreover, Gaussian thermal white noises obeying the
fluctuation-dissipation relation are used; i.e.,
\begin{eqnarray}\label{FDT}
\langle\xi_{1(N)}(t)\rangle &= & 0 \nonumber\\
 \mathrm{and} \hspace{2em}
\langle\xi_{1(N)}(t)\xi_{1(N)}(0)\rangle &= & 2k_B\eta
T_{L(R)}\delta(t) \;.
\end{eqnarray}

Here, $k_B$ denotes the Boltzmann constant and $\eta$ is the coupling
strength between the system and the heat bath.

The time-varying heat bath temperature $T_{L}(t)$ and $T_{R}(t)$ are
chosen as periodic functions $T_{L(R)}(t)=T_{L(R)}(t+2\pi/\omega)$
where $T_0=\overline{T_L(t)}=\overline{T_R(t)}$ is the temporally
averaged, constant environmental reference temperature. Clearly, the coherent changes of these bath temperatures
occur on a time-scale that is much smaller than the (white) thermal fluctuations itself.
Importantly in present context, this so driven system dynamics exhibits a vanishing
average thermal bias; i.e.,
\begin{equation}
\overline{\Delta T(t)}\equiv\overline{T_L(t)-T_R(t)}=0 \;.
\end{equation}

The time-dependent, asymptotic heat flux $J_i(t)$ is assuming the periodicity of the
external driving period $2\pi/\omega$ after the transient behavior
has died out \cite{LHL}. This feature is confirmed in our numerical simulations
(not shown here). At those asymptotic long times, the resulting heat flux
equals the thermal Brownian noise average \cite{HuLiZhao}:
\begin{equation}\label{j-no}
J_i(t)=\left<\dot{x}_{i}\partial{V(x_{i+1},x_i)}/\partial{x_i}\right> \;.
\end{equation}
The stationary heat flux  $\bar{J}$ then follows as the cycle average
over a full temporal period; i.e.,
\begin{eqnarray}\label{j-av}
\overline{J(t)}:= \bar{J}&=&\frac{\omega}{2\pi}
\int_{0}^{2\pi/\omega}
\left<\dot{x}_{i}\partial{V(x_{i+1},x_i)}/\partial{x_i}\right>
dt\nonumber\\ &=&\frac{\omega}{2\pi} \int_{0}^{2\pi/\omega} J_i(t)
dt \;,
\end{eqnarray}
which after averaging becomes independent of atom position $i$.
With ergodicity being obeyed, this double-average equals as well the  long
time average
\begin{align}
\bar{J}&=  \overline{\dot{x}_{i}\partial{V(x_{i+1},x_i)}/\partial{x_i}} \nonumber \\
 &= \lim_{t\rightarrow \infty} \frac{1}{t} \int_{0}^{t} \dot{x}_{i}(t)\partial{V(x_{i+1},x_i)}/\partial{x_i}|_t dt\;.\label{j-over}
\end{align}
Here, the temporal average is over corresponding stochastic trajectories entering the relation in Eq. (\ref{j-av}).
We emphasize  that the resulting
ratcheting heat flux involves an average over the Brownian thermal
noise forces. Put differently, the  flux is not determined by a deterministic  molecular dynamics but
rather by the driven nonlinear Langevin dynamics
following from the  nonlinear lattice dynamics in  Eq. (\ref{nonlinear lattice}) and being
complemented with  Langevin forces obeying the fluctuation-dissipation relation in Eq. (\ref{FDT}).
This in turn defines our Brownian heat motor dynamics.

%Figure 2
\begin{figure}
\includegraphics[width=\columnwidth]{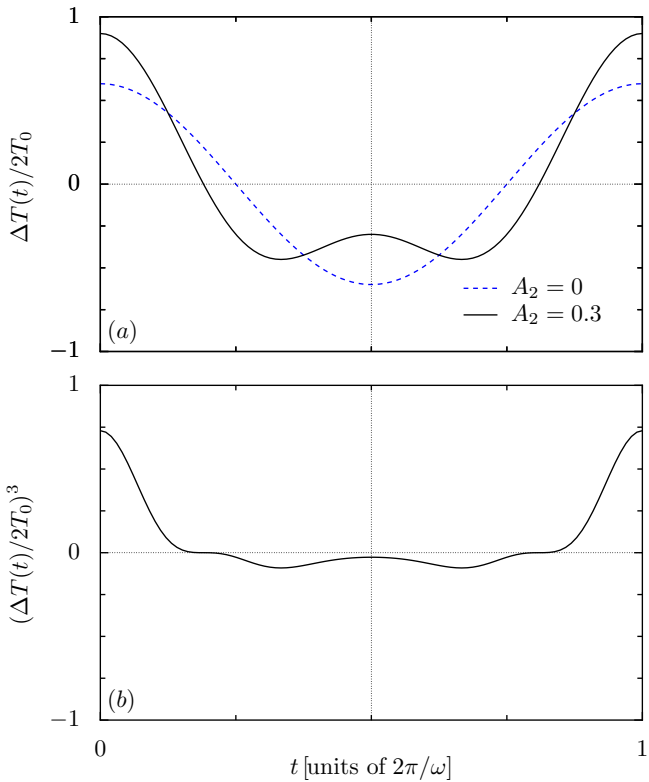}
\vspace{-.5cm} \caption{(Color-online) Modulating the bath
temperature. The temperature bias $\Delta T(t)\equiv T_L(t)-T_R(t)$
is depicted over a full driving period without and with a second
harmonic driving term, panel(a). The strength $A_1=0.6$ is chosen
for both cases. The third moment of the harmonic mixing signal
$(\Delta T(t)/2T_0)^3$ is shown with panel (b). Note that in
distinct contrast to the unbiased first moment the cycle average of
this odd third moment is now nonvanishing.} \label{fig:2}
\end{figure}

\section{ Rocking temperature of one heat bath only}

We start out by considering that only the temperature of the left
heat bath is subjected to a  time-varying modulation, i.e. our case
(i) is defined by setting
\begin{eqnarray}\label{T-field-left}
T_L(t):=T_L&=&T_0[1+A_{1}\cos{(\omega t)}]\nonumber\\
T_R(t):=T_R&=&T_0 \;.
\end{eqnarray}
The symmetric temperature difference $\Delta T(t)/2T_0$ over one period is
depicted in Fig. \ref{fig:2}(a) as the dashed line.

The Fermi-Pasta-Ulam $\beta$ (FPU-$\beta$) lattice is used for
illustration, where the potential term $V(x_{i+1},x_i)$ assumes the
following form:
\begin{equation}\label{v-fpub}
V(x_{i+1},x_i)=\frac{k}{2}(x_{i+1}-x_i-a)^2+\frac{\beta}{4}(x_{i+1}-x_i-a)^4
\end{equation}
We   introduce the dimensionless parameters by measuring positions
in units of $a$, momenta in units of $[a (mk)^{1/2}]$, temperature
in units of $[k a^2/k_B]$, spring constants in units of $k$,
frequencies in units of $[(k/m)^{1/2}]$ and energies in units of $[k
a^2]$. The equations of motion are integrated by the symplectic
velocity Verlet algorithm with a small time step $h=0.005$.  The
system is simulated at least for a total time $t_{tot}=2 \cdot
10^8$. The chosen optimal coupling strength of the heat bath is
fixed at $\eta=0.5$. In the following simulations, we will take the
dimensionless parameters $k=1,\beta=1$.

%Figure 3
\begin{figure}
\includegraphics[width=\columnwidth]{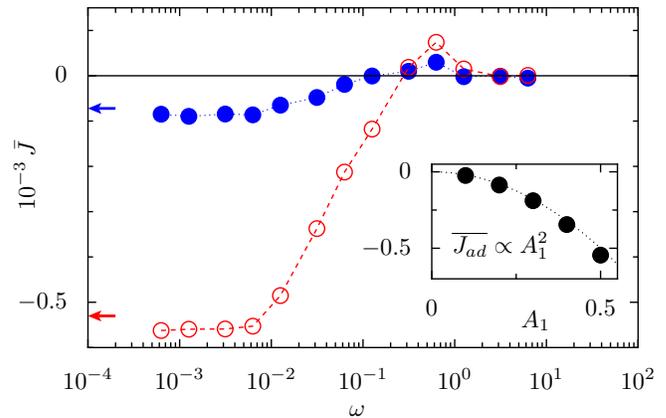}
\vspace{-.5cm} \caption{(Color-online) One-sided temperature
modulation. The cycle averaged ratcheting  heat flux $\bar{J}$ in a
nonlinear FPU-$\beta$ lattice is depicted as the function of the
driving frequency $\omega$. The hollow (red) circles are the results
for harmonic drive $ A_1 \cos(\omega t)$ with $A_1=0.5$; the filled
(blue) circles are the results for a weaker driving strength at
$A_1= 0.2$. The simulations are performed on a FPU-$\beta$ lattice
with $N=50$ atoms and with a reference temperature set at $T_0=0.5$.
The arrows pointing towards the left mark the  adiabatic linear
transport result from Eq. (\ref{j-ad}), see text. Inset is the
figure for $\overline{J_{ad}}$ as the function of $A_1$ and
$\overline{J_{ad}}$ has the basic units $[10^{-3}]$. } \label{fig:3}
\end{figure}

The net heat flux $\bar{J}$ as a function of driving frequency
$\omega$ is depicted in Fig. \ref{fig:3}. In the adiabatic limit
$\omega\rightarrow 0$, a negative ratchet heat flux is obtained. In
the high frequency limit $\omega\rightarrow\infty$, the left- and
right-end atoms will experience a time-averaged constant
temperature. This corresponds to  effective thermal equilibrium,
yielding $\bar{J}\rightarrow 0$ when $\omega\rightarrow\infty$. In
the non-adiabatic limit we even observe a reversal of heat flux,
although not very pronounced.

When the temperature is modulated slowly enough, the periodic
influence of dynamical thermal bias can be viewed as the average,
integrated quasi-stationary flux induced by the momentarily static
thermal bias, i.e. $J_i(t)=J_i(\Delta T(t)/N)$. In the linear
transport regime, this flux can be expressed as a linear transport
Law, reading $J_i(t)=\kappa\left((T_L(t)+T_R))/2\right)\Delta
T(t)/N$. Here the temperature value of the size (N)-dependent
thermal conductivity of the FPU lattice must be determined
numerically at the midpoint of the, in this adiabatic case, linearly
varying temperature profile \cite{EPL_lnb}. Thus,  the adiabatic net
heat flux assumes in leading order the result

\begin{eqnarray}\label{j-ad}
\overline{J_{ad}} &=&\frac{\omega}{2\pi}\int^{2\pi/\omega}_{0}dt\,
\kappa\left(\frac{T_L(t)+T_R}{2}\right)\frac{T_L(t)-T_R}{N} \nonumber\\
&=&\frac{\omega}{2\pi}\int^{2\pi/\omega}_{0}dt\,
\kappa\left(T_0+\frac{A_1 T_0\cos{(\omega
t)}}{2}\right)\frac{A_1 T_0\cos{(\omega t)}}{N}  \nonumber\\
&=&\frac{\omega}{2\pi}\int^{\pi/\omega}_{0}dt\,
\left[\kappa\left(T_0+\frac{A_1 T_0\sin{(\omega
t)}}{2}\right)\right.\nonumber\\
&&\,\,\,\,\,\,\,\,\,\,\,\,\,\,\,\,\,\,\,\,\,\,\,\,\,\,\,\,\,\,
\left.-\kappa\left(T_0-\frac{A_1 T_0\sin{(\omega
t)}}{2}\right)\right] \frac{A_1 T_0\sin{(\omega t)}}{N}\nonumber\\
&<& 0 \;.
\end{eqnarray}
The predicted sign on last line in the above expression originates
from the following reasoning: For the considered FPU-$\beta$ lattice
the thermal conductivity possesses a temperature-dependent behavior
$\kappa(T)\propto 1/T$ in the regime of dimensionless temperature $T
(t)<1$ \cite{EPL_lnb}. With the reference temperature $T_0=0.5$,
$\kappa\left(T_0+\frac{A_1\sin{(\omega t)}}{2}T_0\right)$ is alwaysf
less than $\kappa\left(T_0-\frac{A_1\sin{(\omega t)}}{2}T_0\right)$
in the time window of $[0,\pi/\omega]$. Therefore, a negative
ratchet heat flux will result from the temperature modulation of
only one heat bath as in Eq. (\ref{T-field-left}). This prediction
is corroborated with our numerical calculations in Fig. \ref{fig:3}
for low driving frequencies $\omega$. Furthermore, the adiabatic
value of $\overline{J_{ad}}$ can be approximately calculated from
the quadrature in Eq. (\ref{j-ad}). This calculated adiabatic value
is marked with the arrow pointing towards the left axis in Fig.
\ref{fig:3}. For small rocking strength, $A_1 \ll 1$, this adiabatic
prediction remarkably well agrees with the full nonlinear,
asymptotic value, see in Fig. \ref{fig:3}. For larger driving
strengths $A_1$ the numerically precise full adiabatic result
exhibits notable deviations from this linear transport Law estimate.

% new discussion
Taking into account the size dependent property of heat conductivity
$\kappa(T,N)$, we can express $\overline{J_{ad}}$ further by taking
a Taylor expansion for $\kappa\left(T_0+\frac{A_1\sin{(\omega
t)}}{2}T_0,N\right)$ and $\kappa\left(T_0+\frac{A_1\sin{(\omega
t)}}{2}T_0,N\right)$ at reference temperature $T_0$:
\begin{eqnarray}
\overline{J_{ad}} &=&\frac{\omega}{2\pi}\int^{\pi/\omega}_{0}dt\,
\left[\left.\frac{\partial{\kappa(T,N)}}{\partial{T}}\right|_{T_0}A_1
T_0 \sin{(\omega
t)}+O(A_1^3)\right]\nonumber\\
&&\,\,\,\,\,\,\,\,\,\,\times A_1 T_0 \sin{(\omega t)}/N \nonumber\\
&\approx&\frac{1}{4}A_1^2T_0^2\left.\frac{\partial\kappa(T,N)}{\partial{T}}\right|_{T_0}/N\nonumber\\
&\propto& A_1^2 N^{\alpha-1}
\end{eqnarray}
where in the last line we have used the fact that
$\left.\frac{\partial\kappa(T,N)}{\partial{T}}\right|_{T_0}$ assumes
the same size-dependence as $\kappa(T,N)$, i.e. $\kappa(T,N) \propto
N^{\alpha}$ with $\alpha<1$ \cite{Lepri_pr}. The non-zero adiabatic
ratchet heat flux is a finite size effect since it vanishes in the
limit $N\rightarrow\infty$ proportional to $N^{\alpha-1}$.

We  numerically determined the adiabatic net heat flux by averaging
heat flux from four lowest frequencies for each amplitude $A_1$. The
amplitude effect $\overline{J_{ad}}\propto A_1^2$ is verified by our
numerical calculations as can be seen in the inset of Fig.
\ref{fig:3}.

\section{ Rocking temperature in both heat baths}

We next consider case (ii) with unbiased temperature modulations applied to both heat baths. The
time-varying heat bath temperature $T_{L}(t)$ and $T_{R}(t)$ are
chosen  as:
\begin{eqnarray}\label{T-field}
T_L(t):=T_L=T_0[1+A_{1}\cos{(\omega t)}+A_{2}\cos{(2\omega
t +\varphi)}]\nonumber\\
T_R(t):=T_R=T_0[1-A_{1}\cos{(\omega t)}-A_{2}\cos{(2\omega t
+\varphi)}]
\end{eqnarray}
where $T_0=\overline{T_L(t)}=\overline{T_R(t)}$. The  overall averaged net
temperature difference, $\overline{\Delta
T(t)}\equiv\overline{T_L(t)-T_R(t)}=0$, is again unbiased as before.  The driving amplitude $A_1$ and
$A_2$ are taken as positive values and the relation $A_1+A_2\leq 1$
must be fulfilled to avoid negative temperatures.
Note, however, that in distinct contrast to the situation with case (i) the time-dependent average
temperature; i.e.,

\begin{eqnarray} \label{T-average}
T_{av}(t)\equiv [T_L(t)+T_R(t)]/2=T_0
\end{eqnarray}
is now {\it time-independent}. This feature excludes us from
estimating the adiabatic heat flux  within a linear transport Law of
heat as exercised under case (i). Put differently, the temperature
dependent, nonlinear  thermal conductivity $\kappa(T)$ alone is not
sufficient to set up an adiabatic heat flow, see below. A resulting
finite heat flux is thus beyond the mere role of a nonlinear lattice
and instead is the outcome of the nonlinear interplay of harmonic
mixing of the two frequencies in the nonlinear lattice dynamics to
yield a zero-frequency response for the time-averaged nonlinear heat
flow as defined by Eqs. (\ref{j-av},\ref{j-over}).

In more detail: The second harmonic driving $A_{2}\cos(2\omega t
+\varphi)$ causes nonlinear frequency mixing. The temperature signal
$\Delta T(t)$ notably is unbiased with zero average. It causes,
however, a dynamical symmetry breaking \cite{EPL_luczka,EPL_hanggi};
thus giving rise to directed transport
\cite{EPL_luczka,EPL_hanggi,goychuk,flach,denisov}.
 The time evolution of the harmonic mixing  signal is depicted in Fig.
 \ref{fig:2}(a) for  a phase shift of $\varphi=0$ with the solid line.
The second harmonic drive may  typically contain a non-zero phase shift
 $\varphi$. In accordance with previous studies in single particle
 Brownian motors \cite{goychuk, denisov} the resulting current is expected to become maximal for $\varphi=0$.
 In the following we therefore stick in our numerical studies, if not stated  otherwise,
 to a fixed vanishing phase-shift $\varphi=0$.
Note that this drive with $\varphi=0$ is symmetric under
 time-reversal $t \longrightarrow -t$; nevertheless, time reversal is
 broken by  the frictional Langevin
 dynamics acting upon the end atoms, see above.

 The fact that a
 finite heat flux results when driven by harmonic mixing can be reasoned
 physically by noting that in contrast to the case with $A_2=0$:
 When $A_2 \neq 0$, despite the vanishing temporal cycle average
 $\overline{\Delta T(t)}=
 %\omega/(2\pi) \int_{0}^{2\pi/\omega} \Delta T(t) dt =
 0$, one deals with temporal temperature differences $\Delta T(t)$ that are no longer symmetric
around  $\Delta T(t)= 0$, cf. \ref{fig:2}(a). With this harmonic
mixing modulation, all odd-numbered moments $(\overline{\Delta
T(t))^{(2n+1)}} \neq 0$ , $ n \ge 1 $ are non-vanishing after the
temporal cycle average. With heat flow in nonlinear systems
typically being a function of the temperature bias beyond linear
response regime, we thus nevertheless expect a net finite heat flux.
This feature follows by observing that a leading nonlinear response
due to the non-vanishing temporal cycle average of the third moment
$\overline{(\Delta T(t)/2T_0)^3} =(3/4)A_1^2A_2 \cos{\varphi}$,  see
in Fig. \ref{fig:2}(b), is non-vanishing. In clear contrast to a
single particle case, see in Ref. \cite{goychuk,BM5FN}, the
amplitude $A_2$ in our case (ii) assumes within $T_{L(R)}$ both
signs. Therefore, one principally cannot even predict a priori the
{\it sign} of the resulting heat current. This very fact is the
benchmark of a truly Brownian heat motor where the external
temperature modulation is only weakly coupled to the  Brownian
motion induced heat flow \cite{BM1,BM2,BM3,BM4,BM5}.

\subsection{Shuttling heat across a Fermi-Pasta-Ulam $\beta$ lattice}
\begin{figure}
\includegraphics[width=\columnwidth]{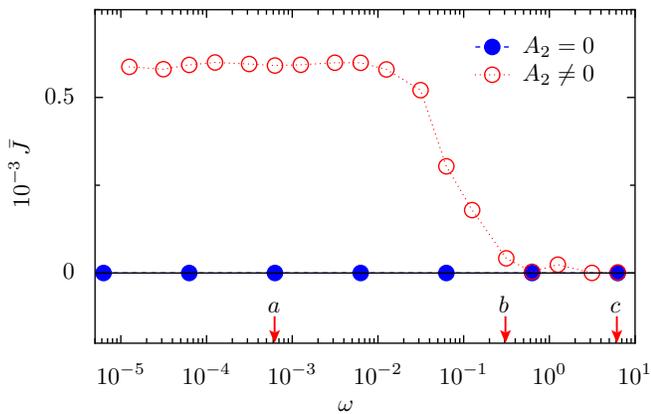}
\vspace{-.5cm} \caption{(Color-online) Ratcheting heat flux in the
FPU-$\beta$-lattice. Cycle averaged heat flux $\bar{J}$ as the
function of the driving frequency $\omega$. The solid circles are
the results for symmetric harmonic driving with $A_2=0$. The hollow
circles are the results for harmonic mixing with $A_2=0.3$ and a
relative phase shift $\varphi =0$. The harmonic driving parameter is
set for both cases at $A_1=0.6$. The simulations are performed on a
FPU-$\beta$ lattice with $N=50$ atoms and with a reference
temperature $T_0=0.5$. The three arrows indicate the distinct
driving frequencies used in Fig. \ref{fig:5}.} \label{fig:4}
\end{figure}

We start the study of case (ii) by considering  the FPU-$\beta$ lattice of Eq. (\ref{v-fpub}) by using the same parameters as before.
The net heat flux $\bar{J}$
as a function of driving frequency $\omega$ is depicted in Fig.
\ref{fig:4}. For $A_2=0$, there indeed emerges no finite heat flux; this corroborates with theory because of the
absence of dynamical symmetry breaking between positive and negative
temperature differences $\Delta T(t)$, see Fig.
\ref{fig:2}. The non-zero second harmonic driving term with $A_2 \neq
0$ globally causes with $\Delta T(t) \neq 0$  a dynamical symmetry breaking. At low, adiabatic
driving frequencies $\omega\rightarrow 0$, we  obtain a finite heat
flux $\bar{J}\neq 0$ which is solely induced by dynamical symmetry
breaking caused by harmonic mixing. Again, the heat flux $\bar{J}$ vanishes
in the high frequency limit as expected.

\subsubsection*{Adiabatic estimate}
An adiabatic analysis within a linear transport mechanism is no
longer possible here: By noting that the  average temperature
$[T_L(t)+T_R(t)]/2=T_0$ is {\it time-independent}, the previous
adiabatic estimate now reduces to:
\begin{eqnarray}
\overline{J_{av}}&=&\frac{\omega}{2\pi}\int^{2\pi/\omega}_{0}dt\,
\kappa\left(\frac{T_L(t)+T_R(t)}{2}\right)\frac{T_L(t)-T_R(t)}{N}\nonumber\\
&=&\frac{\omega}{2\pi}\int^{2\pi/\omega}_{0}dt\, \kappa\left(T_0\right)\frac{\Delta T(t)}{N}\nonumber\\
&=&0 \,;
\end{eqnarray}
i.e., it predicts a vanishing rachet heat flux. As mentioned above, the observed finite heat flux
is due to a nonlinear interplay between
the driving frequencies for which we are even not able to predict {\it a priori} the sign.

%Figure 4

%Figure 5
\begin{figure}
\includegraphics[width=\columnwidth]{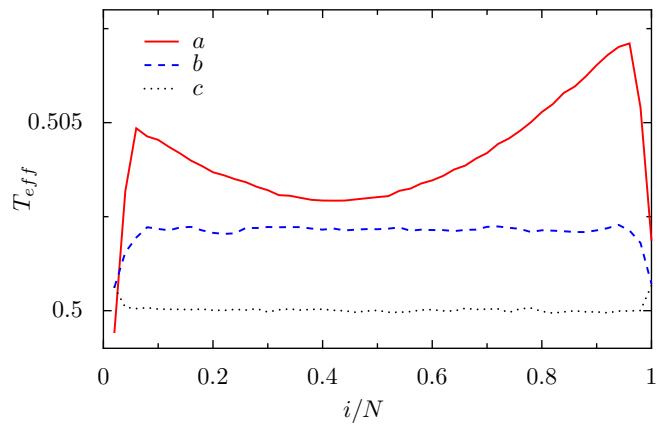}
\vspace{-.5cm} \caption{(Color-online) Effective temperature
profiles of three selected frequencies for $A_2=0.3$ in Fig.
\ref{fig:4} with $T_0=0.5$.} \label{fig:5}
\end{figure}

\subsubsection*{Local temperature profile}
To gain further insight into this intriguing regime of finite
ratchet heat flux, we investigate the local temperature variations
across the chain at three distinct different driving frequencies
$\omega$. The local (kinetic) temperature is defined via the
equipartition theorem as the time average of kinetic energy
$T_{eff}(i)=\overline{\dot{q}^2_i}$, as in Ref. \cite{LHL}. The
effective temperature profiles of three numerical runs, denoted as
($a, b, c$) in Fig. \ref{fig:4}, are depicted in Fig. \ref{fig:5}
versus the relative site position $i/N$. In clear contrast to the
non-driven case with no temperature modulation, a distinct
temperature profile emerges for the driven case. This averaged
effective temperature $T_{eff}(i)$ lies typically above the
time-independent  average temperature $T_{av} = T_0$ with regimes of
both, positive- and negative-valued local gradient. Even for the
case of slow, adiabatic driving, i.e., case (a) in Fig. \ref{fig:5},
we cannot now even detect a clear-cut mechanism to yield the now
positive sign for the shuttled heat flux. This corroborates with our
reasoning that  no simple adiabatic estimate can be devised in this
situation.

We also note that with identical signs for $A_1,A_2$ of
$T_{L(R)}(t)$ in Eq. (\ref{T-field}), implying that $\Delta T(t)=0$
identically, no ratchet heat flux can be detected (not shown).

%We find that in the adiabatic limit, the inner structure of the
%effective temperature profile exhibits a relationship with the observed direction
%of the  heat flux. For example,  for the driving frequency at the value (a), the inner temperature profile
%can be viewed as the combination of two effective temperature
%gradients with opposite directions. The positive effective
%temperature gradient profile assumes a lower average temperature than the
%negative effective temperature gradient. Using the fact that in this case, too,
%$\kappa(T)\propto 1/T$ \cite{EPL_lnb}, this intrinsic heat transport structure indicates a positive
%ratchet heat flux,  in agreement with our numerical findings. Finally, we also note that with identical signs for $A_2$ in
%$T_{L(R)}$, implying that $\Delta T(t) = 0$ identically, we detect zero ratchet heat flux (not shown).

\subsection{Shuttling heat across a Fermi-Pasta-Ulam $\alpha\beta$ lattice}

We next consider the Fermi-Pasta-Ulam $\alpha\beta$
(FPU-$\alpha\beta$) lattice, where the potential term
$V(x_{i+1},x_i)$ assumes the following form:
\begin{eqnarray}\label{v-fpu}
&&V(x_{i+1},x_i)=\frac{k}{2}(x_{i+1}-x_i-a)^2\nonumber\\
&&\,\,\,\,+\frac{\alpha}{3}(x_{i+1}-x_i-a)^3+\frac{\beta}{4}(x_{i+1}-x_i-a)^4
\end{eqnarray}
In the  simulations, we will employ the dimensionless
parameters $k=1,\alpha=1,\beta=1/4$.

\begin{figure}
\includegraphics[width=\columnwidth]{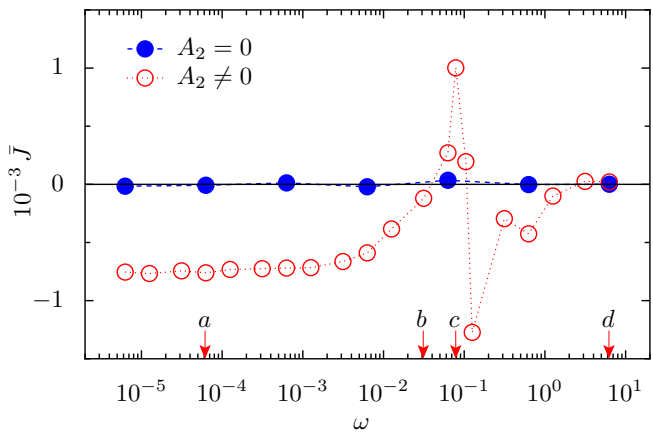}
\vspace{-.5cm} \caption{(Color-online) Ratcheting heat flux in a
FPU-$\alpha\beta$ lattice. Cycle averaged heat flux $\bar{J}$ as the
function of the driving frequency $\omega$. The solid circles are
the results for symmetric harmonic driving with $A_2=0$. The hollow
circles are the results for harmonic mixing with $A_2=0.3$ and a
relative phase shift $\varphi =0$. The harmonic driving parameter is
set for both cases at $A_1=0.6$. The simulations are performed on a
FPU-$\alpha\beta$ lattice with $N=50$ atoms and with a reference
temperature $T_0=1$, see text. The arrows indicate the driving
frequencies used in Fig. \ref{fig:7}.} \label{fig:6}
\end{figure}

The net heat flux $\bar{J}$ for this system is depicted in Fig.
\ref{fig:6}.  We again cannot detect any finite heat flux for
$A_2=0$. For $A_2=0.3$, we now detect a negative ratchet heat flux
in the low frequency, adiabatic limit. This ratchet heat flux
vanishes as expected in the high frequency limit. Surprisingly
however, the heat flux does not approach zero monotonically as in
the case with the FPU-$\beta$ lattice. At some intermediate
frequency range, the direction of the net heat flux reverses sign!
We detect two distinct peaks for the heat flux of {\it opposite}
directions that occur within a narrow frequency window.

%Figure 6

%Figure 7
\begin{figure}
\includegraphics[width=\columnwidth]{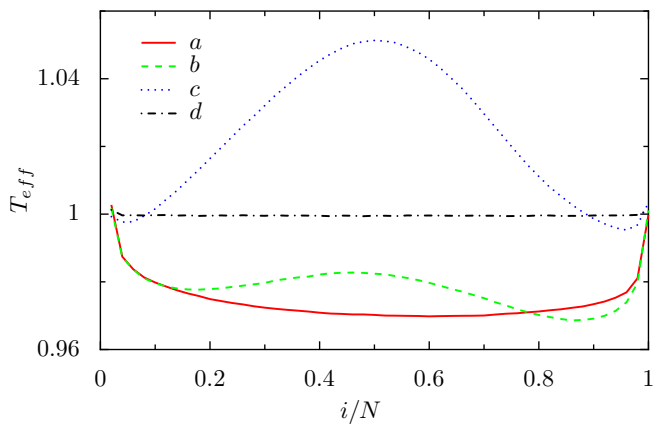}
\vspace{-.5cm} \caption{(Color-online) Effective temperature
profiles for the four selected frequencies with $A_2=0.3$ depicted in Fig.
\ref{fig:6}. Here the average bath temperature is held at $T_0=1$.} \label{fig:7}
\end{figure}

We as well depict a plot for the local temperature variations across
the chain at different driving frequencies $\omega$. The effective
temperature profiles of four numerical runs, denoted as ($a, b, c,
d$) in Fig. \ref{fig:6}, are shown in Fig. \ref{fig:7} versus the
relative site positions $i/N$. For frequencies near the
flux-reversal point, the inner structure of effective temperature
becomes very complicated and we could not detect a clear-cut
connection between local temperature and the direction of ratcheting
heat flux.

\begin{figure}
\includegraphics[width=\columnwidth]{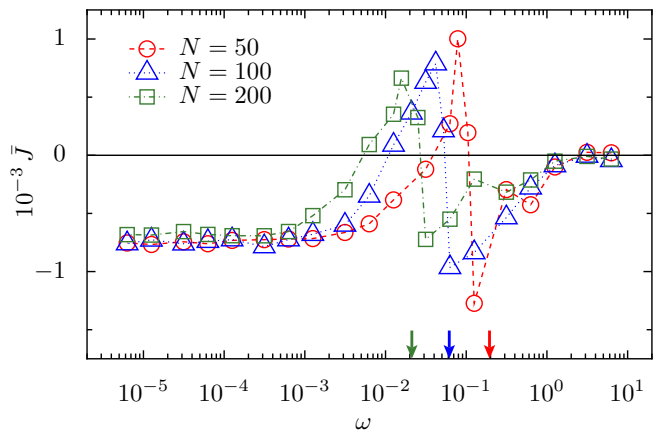}
\vspace{-.5cm} \caption{(Color-online) Size dependence of directed
heat flux. Ratchet heat flux $\bar{J}$ {\it vs.} the driving
frequency $\omega$ for different FPU-$\alpha\beta$ lattice sizes
$N=50$ (circle), $N=100$ (triangle) and $N=200$ (square). The
driving amplitudes are set at $A_1=0.6$,  $A_2=0.3$ and $\varphi=0$.
The estimate of the frequency scale set by the thermal response time
which scales with the anomalous FPU-heat conductivity $\kappa(N)$.}
\label{fig:8}
\end{figure}

When the lattice size $N$ is increased, the thermal bias $\Delta
T(t)/N$ is reduced. One therefore would expect a  smaller heat flux
$\bar{J}$. In Fig. \ref{fig:8}, we plot  $\bar{J}$ versus $\omega$
for three different lattice sizes $N=50, 100, 200$. Contrary to
common intuition, however, the finite heat flux $\bar{J}$ in the
adiabatic limit $\omega\rightarrow 0$ is practically {\it
independent} of system size. This feature again corroborates with
the fact that in this regime no obvious adiabatic estimate is
deducible.

This result of a size-insensitivity will prove advantageous for the
experimenters attempting to measure such directed ratchet heat flux,
e.g. by use of a nanotube coupled in between two heat contacts.

The peaks  of opposite directions are also found for larger system
sizes. The positions of these peaks become red-shifted for larger
system size $N$. This effect is related to the thermal response time
as we have detailed before with  Ref. \cite{LHL}. The characteristic
frequency can be estimated as $\omega_c\approx 8\pi\kappa(N)/cN^2$.
The specific heat can be approximated as $c\approx 1$. The FPU
lattice is well known to exhibit anomalous heat conduction with
size-dependent thermal conductivities $\kappa(N)$
\cite{kabu,lepri,pere}. The thermal conductivity $\kappa(N)$ at
temperature $T=1$ has the following numerical values,
$\kappa(50)=19.1, \kappa(100)=25.1$ and $\kappa(200)=32.9$. Thus,
the characteristic frequencies can be estimated as $\omega_c\approx
0.19$ for $N=50$, $\omega_c\approx 0.063$ for $N=100$ and
$\omega_c\approx 0.021$ for $N=200$. These values are marked as
arrows pointing to $x$-axis in Fig. \ref{fig:8}; these values
impressively corroborate with  numerical simulation results.

%Figure 8

\subsection {Directing heat across a Lennard-Jones lattice}

We finally consider the physically  realistic case of a
Lennard-Jones (LJ) lattice. The interaction $V(x_{i+1},x_i)$ for a
LJ lattice interaction takes the form
\begin{equation}
V(x_{i+1},x_i)=\epsilon\left[\left(\frac{a}{x_{i+1}-x_i}\right)^{12}
-2\left(\frac{a}{x_{i+1}-x_i}\right)^{6}\right] \;,
\end{equation}
where $a$ is the lattice constant and $\epsilon$ is the depth of the
potential well. New dimensionless parameters can be introduced by
measuring positions in units of $a$, momenta in units of
$[(m\epsilon)^{1/2}]$, temperature in units of $[\epsilon/k_B]$,
spring constants in units of $[\epsilon/a^2]$, frequencies in units
of $[(\epsilon/ma^2)^{1/2}]$ and energies in units of $\epsilon$. A
particular material is described by the pair of parameters $a$ and
$\epsilon$.
%, the intensity reflecting the spacial symmetry
%breaking can only be determined by the temperatures.

\begin{figure}
\includegraphics[width=\columnwidth]{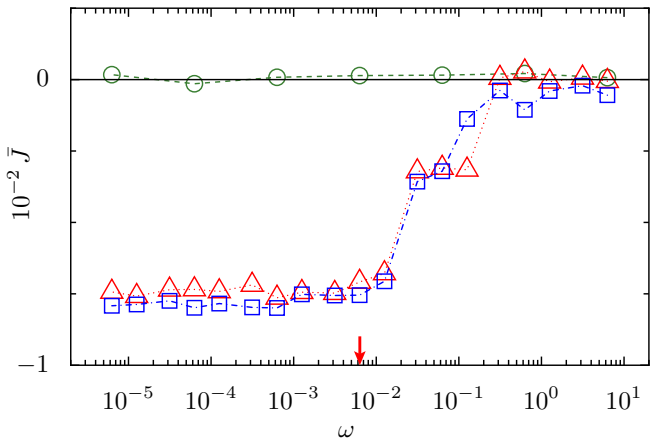}
\vspace{-.5cm} \caption{(Color-online) Ratcheting heat flux across a
Lennard-Jones lattice. Cycle averaged heat flux $\bar{J}$ {\it vs.}
the harmonic mixing driving frequency $\omega$ for a LJ-lattice.
Circles are the results for the lattice size $N=50$ in absence of
second harmonic drive, i.e. $A_2=0$. Triangles and squares are the
results for $N=50$ and $N=100$ with the second harmonic driving set
at $A_2=0.3$. The fundamental driving amplitude  is set at $A_1=0.6$
for all cases. The reference temperature is chosen as $T_0 = 3$
which corresponds to $\approx 360 $\/[K]. The arrow indicates the
driving frequency used in Fig. \ref{fig:8}.} \label{fig:9}
\end{figure}

The resulting ratchet heat flux $\bar{J}$ for LJ-lattice is depicted
in Fig. \ref{fig:9}. Just as is the case in FPU lattice, we cannot
detect finite flux $\bar{J}$ for pure harmonic driving with $A_2=0$.
In the adiabatic limit $\omega\rightarrow 0$, a virtually {\it
size-independent} finite heat flux $\bar{J}$ results when $A_2 \neq
0$; i.e. the emerging ratchet heat flux is rather robust. In our
simulations we used a dimensionless reference temperature $T_0=3$.
To give a example, for Argon atoms with parameters $(a=3.4
[\mathring{A}],\epsilon=119.8 k_{B}[K])$, this dimensionless
temperature $T_0$ corresponds to a physical temperature
$T=T_{0}\epsilon/k_B=359.4 [K]$.

%Figure 9

%Figure 10
\begin{figure}
\includegraphics[width=\columnwidth]{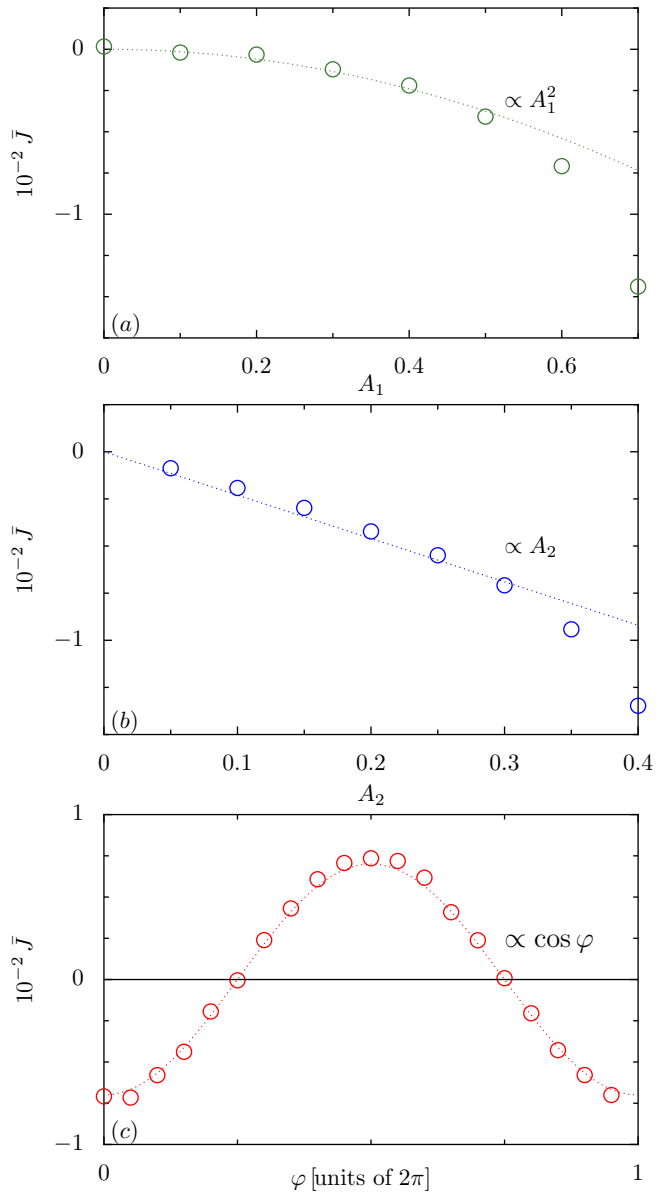}
\vspace{-.5cm} \caption{(Color-online) Ratcheting heat flux across a
Lennard-Jones lattice. (a) Cycle averaged heat flux $\bar{J}$ {\it
vs.} the fundamental driving amplitude $A_1$ with fixed $A_2=0.3$
and $\varphi=0$. (b) Cycle averaged heat flux $\bar{J}$ {\it vs.}
the driving amplitude $A_2$ with fixed $A_1=0.6$ and $\varphi=0$.
(c) Cycle averaged heat flux $\bar{J}$ {\it vs.} the phase shift
$\varphi$ with fixed $A_1=0.6$ and $A_2=0.3$. For all cases, the
lattice size is set as $N=50$ and the reference temperature is
chosen as $T_0 = 3$. } \label{fig:10}
\end{figure}

\subsubsection*{Controlling heat flux via the phase shift $\varphi$}

For single particle Brownian motors, it is well known that the
directed transport in leading order of nonlinearity is proportional
to the non-vanishing time-averaged third moment \cite{goychuk,
denisov}. Whether this result still holds true for a spatially
extended ratchet system can be tested in the present context
numerically only. We conjecture that the directed current still will
be proportional to the third order moment $\overline{(\Delta
T(t)/2T_0)^3}$, i.e. $\bar{J}\propto A_1^2A_2\cos{\varphi}$. The
numerically evaluated  heat flux $\bar{J}$ is depicted {\it vs.}
$A_1$, $A_2$ and $\varphi$ in  Fig. \ref{fig:10}. As expected from
theory, it is only for small values of driving amplitudes $A_1$
($A_1<0.5$ in panel (a)) and $A_2$ ($A_2<0.3$ in panel (b)) that the
heat current response $\bar{J}$ follows the theoretical scaling law
$\bar{J}\propto A^2_1A_2$ with good accuracy. Note that the sign of
the flux remains, however, undetermined from the form of harmonic
mixing with opposite signs for $ \pm A_2$. At larger driving
strengths, higher order nonlinear contributions yield a sizable
contribution, thus causing deviations from the leading scaling
behavior. Interestingly enough, however, the heat current $\bar{J}$
is found to exhibit the  dependence on the relative phase shift
$\bar{J}\propto \cos{\varphi}$ very accurately, even for substantial
large driving strengths; i.e. for $A_1 =0.6$ see in Fig.
\ref{fig:10}(c).  This numerical finding is advantageous for the
control of heat current in spatially extended systems: the direction
of directed heat flow can be reversed by merely adjusting the
relative phase shift $\varphi$ of the second harmonic drive.

\section{ Conclusions}
In conclusion, we have demonstrated sizable  shuttling of  a net heat
flux across 1D homogenous nonlinear lattices of the FPU-type and Lennard-Jones type by applying two
different symmetry breaking mechanisms. These different mechanisms are imposed via  temporal modulations of
the bath temperature(s). For a modulation of temperature of one heat bath only
 a  symmetric harmonic driving is sufficient to induce a
non-vanishing heat flux. The resulting ratchet heat flux can be
elucidated at low driving frequencies by virtue of an adiabatic
estimate in terms of a single quadrature, see Eq. (\ref{j-ad}). The
expression involves the knowledge of the nonlinear temperature
dependent heat conductivity. According to this adiabatic analysis,
in the linear transport regime, the ratchet heat flux is found to be
proportional to the square of driving amplitude and does vanish in
the thermodynamic limit $N\rightarrow\infty$.

For  a situation with temperature modulations applied at both heat
baths, a simple  harmonic driving no longer  induces a ratchet heat
flux. If a second harmonic driving is included one can  break the
symmetry dynamically. The resulting ratchet heat flux obeys,
however, no simple adiabatic estimate so that even the direction of
the ratcheting heat flux cannot be predicted a priori. Moreover, we
detect in the non-adiabatic regime a distinct reversal of heat flux
for the case of the FPU-$\alpha\beta$ chain: It occurs around the
thermal response time set by the anomalous heat conductivity. For
the realistic situation with a Lennard-Jones lattice we find that
the resulting flux of the Brownian heat motor is robust and is
practically independent of system size.

Noteworthy is the fact that the directed
heat flux is  substantially larger for a physically realistic  Lennard-Jones lattice as compared to the situation
of coupled, non-identical Frenkel-Kontorova
lattices, see in Ref.  \cite{LHL}. As a consequence, an  experimental setup
as put forward with this work seems more feasible to realize such a Brownian motor for shuttling heat
as compared to a physical situation with two coupled Frenkel-Kontorova lattices.\\

\acknowledgments The authors like to thank Dr. S. Denisov for his
insightful comments on this work. The work is supported in part by
an ARF grant, R-144-000-203-112, from the Ministry of Education of
the Republic of Singapore, grant R-144-000-222-646 from NUS, by the
German Excellence Initiative via the \textit {Nanosystems Initiative
Munich} (NIM) (P.H.), the German-Israel-Foundation (GIF) (N.L.,
P.H.) and the DFG-SPP program DFG-1243 -- quantum transport at the
molecular scale (F.Z., P.H., S.K.).


\begin{thebibliography}{01}

\bibitem{WangLi08} L. Wang and B. Li, Physics World {\bf 21} (3),
28 (2008).

\bibitem{rectifier} M. Terraneo, M. Peyrard, and G. Casati, Phys. Rev. Lett. \textbf{88}, 094302 (2002).

\bibitem{diode} B. Li, L. Wang, and G. Casati, Phys. Rev. Lett. \textbf{93}, 184301 (2004).

\bibitem{quantumdiode} D. Segal and A. Nitzan, Phys. Rev. Lett. \textbf{94}, 034301 (2005).

\bibitem{Hu06}B. Hu, L. Yang, and Y. Zhang, Phys. Rev. Lett. \textbf{97}, 124302
(2006).
%%%%%%%%%%new
\bibitem{peyrard}
M.~Peyrard, Europhys. Lett. \textbf{76}, 49 (2006).
%%%%%%%%%%%%%%%

\bibitem{yang}N. Yang, N. Li, L. Wang, and B. Li, Phys. Rev. B {\bf 76}, 020301(R) (2007).

\bibitem{transistor} B. Li, L. Wang, and G. Casati, Appl. Phys. Lett. \textbf{88}, 143501 (2006).

\bibitem{WangLi07} L. Wang and B. Li, Phys. Rev. Lett. {\bf 99},
177208 (2007).

\bibitem{wangli08}
L.~Wang and B.~Li, Phys. Rev. Lett. {\bf 101}, 267203 (2008).

\bibitem{experimentaldiode} C. W. Chang, D. Okawa, A. Majumdar, and A. Zettl, Science \textbf{314}, 1121 (2006).

\bibitem{BM1}
P. Reimann, R. Bartussek, R. H\"aussler, and P. H\"anggi, Phys.
Lett. A {\bf 215}, 26 (1996).
\bibitem{BM2}
R. D. Astumian  and P. H\"anggi, Phys. Today {\bf 55} (11), 33 (2002).
\bibitem{BM3}
P. H\"anggi, F. Marchesoni, and F. Nori, Ann. Phys. (Leipzig) {\bf 14},
51 (2005).
\bibitem{BM4}
P. Reimann and P. H\"anggi, Appl. Phys. A \textbf{75},
169 (2002).
\bibitem{BM5}
P.~H\"anggi and F.~Marchesoni, Rev. Mod. Phys. {\bf 81}, 387 (2009).

\bibitem{LHL} N. Li, P. H\"anggi, and B. Li, EPL \textbf{84}, 40009
(2008).

%%%%% new
\bibitem{vandenbroeckPRL2006}
C. Van den Broeck and R. Kawai, Phys. Rev. Lett. \textbf{96}, 210601
(2006).
\bibitem{segal2003}
D.~Segal, A.~Nitzan and P.~H\"anggi, J.~Chem.~Phys. {\bf 119}, 6840 (2003).
\bibitem{segalPRE2006}
D. Segal and A. Nitzan, Phys. Rev. E {\bf 73}, 026109 (2006).
\bibitem{marathePRE2007}
R. Marathe, A. M. Jayannavar and A. Dhar, Phys. Rev. E {\bf 75},
030103(R) (2007).
\bibitem{vandenbroeckPRL2008}
M. van den Broek and C. Van den Broeck, Phys. Rev. Lett. {\bf 100},
130601 (2008).

\bibitem{HuLiZhao}
B. Hu, B. Li, and H. Zhao, Phys. Rev. E \textbf{57}, 2992 (1998).

\bibitem{EPL_lnb}
N. Li and B. Li, EPL {\bf 78}, 34001 (2007).

\bibitem{Lepri_pr}
S. Lepri, R. Livi, and A. Politi, Phys. Rep. {\bf 377}, 1 (2003).

\bibitem{EPL_luczka}
J. ~Luczka, R.~Bartussek and P.~H\"anggi, Europhys.~Lett. {\bf 31},
431 (1995).

\bibitem{EPL_hanggi}
P.~H\"anggi, R. Bartussek, P.~Talkner, and J.~Luczka,
Europhys.~Lett. {\bf 35}, 315 (1996).

\bibitem{goychuk}
I.~Goychuk and P.~H\"anggi, Europhys.~Lett. {\bf 43}, 503 (1998).

\bibitem{flach}
S.~ Flach, O.~Yevtushenko, and Y.~Zolotaryuk, Phys. Rev. Lett. {\bf
84}, 2358 (2000).

\bibitem{denisov}
S.~Denisov, S.~Flach, A.A.~Ovchinnikov, O.~ Yevtushenko, and Y.~
Zolotaryuk, Phys. Rev. E {\bf 66}, 041104 (2002).

\bibitem{BM5FN}
P.~H\"anggi and F.~Marchesoni, Rev. Mod. Phys. {\bf 81}, 387 (2009); see Section II.C.1 therein.



%%%%%%%%%%%%%%%%%%%%%%%%



\bibitem{kabu}
H.~Kaburaki and M.~ Machida, Phys.~Lett. A {\bf 181}, 85 (1993).

\bibitem{lepri}
S.~Lepri, R.~Livi, A.~Politi, Phys. Rev.~Lett. {\bf 78}, 1896
(1997).

\bibitem{pere}
A.~Pereverzev, Phys.~Rev.~E {\bf  68}, 056124 (2003).

\end{thebibliography}
\end{document}